 \documentclass[final,5p,times,twocolumn]{elsarticle}

\usepackage{amssymb}
\usepackage{amsthm}

\usepackage{lineno,hyperref}
\usepackage{cleveref}
\usepackage{mathptmx}
\journal{Nuclear Instruments and Methods in Physics Research A}

\begin{document}

\begin{frontmatter}



\title{EuPRAXIA$@$SPARC$\_$LAB: the high-brightness RF photo-injector layout proposal}

\author[label1]{ A. Giribono }\fnref{myfootnote}
\author[label3]{ A. Bacci}
\author[label1]{ E. Chiadroni }
\author[label2]{ A. Cianchi }
\author[label1]{ M. Croia }
\author[label1]{ M. Ferrario }
\author[label1]{ A. Marocchino }
\author[label3]{ V. Petrillo }
\author[label1]{ R. Pompili }
\author[label1]{ S. Romeo}
\author[label3]{ M. Rossetti Conti}
\author[label3]{ A.R. Rossi}
\author[label1]{ C. Vaccarezza}


\address[label1]{INFN-LNF, Via Enrico Fermi 40, 00044 Frascati Rome, Italy}
\address[label2]{INFN-Tor Vergata University, Via Ricerca Scientifica 1, 00133 Rome, Italy}
\address[label3]{INFN-MI, Via Celoria 16, 20133 Milan, Italy}

\fntext[myfootnote]{Email address: anna.giribono@lnf.infn.it}




\begin{abstract}
At EuPRAXIA$@$SPARC$\_$LAB, the unique combination of an advanced high-brightness RF injector and a plasma-based accelerator will drive a new multi-disciplinary user-facility. The facility, that is currently under study at INFN-LNF Laboratories (Frascati, Italy) in synergy with the EuPRAXIA collaboration, will operate the plasma-based accelerator in the external injection configuration. Since in this configuration the stability and reproducibility of the acceleration process in the plasma stage is strongly influenced by the RF-generated electron beam, the main challenge for the RF injector design is related to generating and handling high quality electron beams. In the last decades of R$\&$D activity, the crucial role of high-brightness RF photo-injectors in the fields of radiation generation and advanced acceleration schemes has been largely established, making them effective candidates to drive plasma-based accelerators as pilots for user facilities. An RF injector consisting in a high-brightness S-band photo-injector followed by an advanced X-band linac has been proposed for the EuPRAXIA$@$SPARC$\_$LAB project. The electron beam dynamics in the photo-injector has been explored by means of
simulations, resulting in high-brightness, ultra-short bunches with up to 3 kA peak current at the entrance of the advanced X-band linac booster. The EuPRAXIA$@$SPARC$\_$LAB high-brightness photo-injector is described here together with performance optimisation and sensitivity studies aiming to actual check the robustness and reliability of the desired working point. 
\end{abstract}


\end{frontmatter}

\begin{figure*}[!ht]
	\begin{center}
		\includegraphics[width=0.9\textwidth]{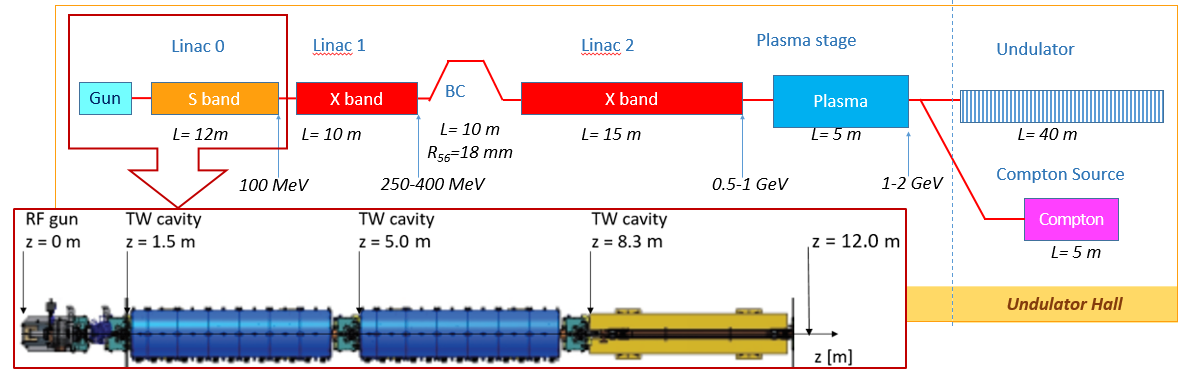}    
		\caption{Layout of the EuPRAXIA@SPARC$\_$LAB RF injector consisting in a SPARC-like high brightness S-band photo-injector coupled with an X-band linac booster. The RF injector is followed by a plasma accelerating stage that will drive user applications. The highlight shows the photo-injector layout consisting of 1.6 cell UCLA/BNL type SW RF gun, equipped with a copper photo cathode and an emittance compensation solenoid, followed by three TW SLAC type  sections \cite{item14}; other two compensation solenoids surround the first and the second S-band cavities for the operation in the velocity bunching scheme.}
		\label{fig:sparc_ph}
	\end{center}
\end{figure*}

\section{Introduction}

The EuPRAXIA@SPARC$\_$LAB project \cite{item308} aims to design and build a compact FEL source, equipped with user beam-lines at 3 nm wavelength, driven by a high gradient plasma-based accelerator. The project arises from the will to candidate the INFN-LNF Laboratories (Frascati, Italy) as host for the EuPRAXIA European Facility \cite{item300}. At EuPRAXIA@SPARC$\_$LAB the electron beam is generated in an advanced compact RF injector coupled with a plasma-based accelerating stage operating in the external injection configuration. The main challenge for the RF injector comes from the request of producing ultra-short ($<$10 fs), high quality electron beams useful to cover a wide range of user applications, as fundamental physics oriented research and high social impact applications. The proposal for the RF injector foresees a SPARC-like S-band photo-injector  \cite{item29} coupled with a X-band linac booster \cite{item301,item302} (see \figurename~\ref{fig:sparc_ph}). The choice for the RF photo-injector has been guided by the expertise acquired at SPARC$\_$LAB \cite{item27} in the stable and routinely generation and manipulation of high brightness electron beams useful for plasma-based experiments \cite{item306,item307} and generation of advanced radiation sources \cite{item41,item22,item37}.

Beam dynamics in the EuPRAXIA@SPARC$\_$LAB accelerator has been studied by means of start to end simulations from the cathode up to the undulator entrance. Four working points have been investigated to provide a final electron beam that fulfils the EuPRAXIA goals, that means at least 1 GeV energy, RMS transverse normalised emittance lower than 1 mm mrad and a 2-3 kA peak current (total charge divided by FWHM bunch length). Two different longitudinal compression schemes are used to satisfy the request on the peak current: a full RF compression in the photo-injector, WP1 and WP4, and a hybrid scheme, coupling the RF compression in the photo-injector and a magnetic chicane in the X-band linac, WP2 and WP3. Once the beam is generated and compressed in the photo-injector, it enters the X-band linac with energy of about 100 MeV (full RF compression) or 170 MeV (hybrid scheme) depending on the adopted longitudinal compression scheme (see \tablename~\ref{tab:table_inj}). 

The nominal Working Points (WPs) are described in the following
\begin{itemize}
	\item WP1, Low Charge-High Current: 30 pC charge in 10 fs FWHM bunch length, which turns into 3 kA peak current suitable both for particle beam and laser driven wakefield acceleration (PWFA and LWFA),
	\item WP2, Low Charge-Low Current: 30 pC charge in 120 fs RMS bunch length, which turns into 0.1 kA peak current suitable for PWFA and LWFA; the bunch is further compressed in a magnetic longitudinal compression stage to reach the desired 3 kA peak current at the plasma entrance,
	\item WP3, high charge-Low Current: 200 pC charge in 400 fs RMS bunch length, which turns into 0.07 kA RMS current, with and without the longitudinal bunch compression in the magnetic chicane, to serve both the SASE-FEL, with peak current I$_{peak}$ = 2 kA, and the Compton and THz sources in the high flux operation scheme.
	\item WP4, Comb operation: Low Charge-High Current trailing bunch following a High Charge driver bunch suitable for PWFA. 
\end{itemize}

\begin{table}[!h]
	\small
	\centering
	\caption{ Electron beam parameters at photo-injector exit resulting from beam dynamics studies.}
	\vskip 0.1 in
	\begin{tabular}{|p{1.8cm}|c|c|c|c|c|} \hline
		Compression &Hybrid  & \multicolumn{4}{|c|}{RF}\\
		\hline
		Operation& \multicolumn{3}{|c|}{Single Bunch} & \multicolumn{2}{|c|}{Comb }\\
		\hline
		&WP2	&WP3	&WP1	&Witness&Driver	\\
		\hline
		Q [pC]								& 30	& 200	& 30	& 30	&200	\\
		\hline
		E [MeV]   							&171.1  &171.4 	& 98.8	& 101.5	&103.0	\\
		\hline
		$\frac{\Delta\gamma}{\gamma}$ [\%] 	&  0.22	& 0.67	& 0.30  &0.15 	&0.67 	\\
		\hline
		$\sigma_{x,y}$  $[\mu m]$ 		&104.0  & 390.0 & 58.4  & 182.0 	&102.0 	\\
		\hline
		$\epsilon_{n_{x,y}}$[mm.mrad]&  0.33	& 0.37	& 0.58  &0.69 	&1.95	\\
		\hline
		$\sigma_{z}$  $[\mu m]$  		& 37	& 112	& 5.6   &3.5	&55.5	\\
		\hline
		$I_{peak} $ [kA] 				& 0.09 	& 0.35  	& 4.00   &6.00  	& 0.37 	\\
		\hline
	\end{tabular}
	\label{tab:table_inj}
\end{table}

\section{The EuPRAXIA@SPARC$\_$LAB photo-injector}

The EuPRAXIA at SPARC$\_$LAB photo-injector, operating at 2.856 GHz, consists of a 1.6 cell UCLA/BNL type SW RF gun followed by three TW SLAC type sections \cite{item14}; the RF gun is equipped with a copper photo cathode and an emittance compensation solenoid while the first and second S-band cavities are surrounded by solenoids. The whole system layout is reported in \figurename~\ref{fig:sparc_ph}.

The gun operates with a peak field at the cathode of $E_{acc}\simeq$ 120 MV/m; a slightly dephasing between the field and the beam allows to maximise the energy gain in this part. The first two S-band structures operate at $E_{acc}$ = 20.0 MV/m and the third one at $E_{acc}$ = 28.0 MV/m on average. The magnetic field of the solenoids surrounding the RF cavities, is such to permits operating the first and second TW sections far from the crest in the velocity bunching regime to enable the bunch longitudinal compression. The third section operates almost on crest in order to let the electron bunch gain the energy and freeze its phase space quality.

The beam dynamics in the EuPRAXIA@SPARC$\_$LAB high-brightness photo-injector has been explored by means of simulations performed with TStep \cite{item6a}. This multi-particle code takes into account the space charge effects, relevant at very low energies, and the thermal emittance. The WPs described above and \tablename~\ref{tab:table_inj} have been investigated.

\section{Low charge, single bunch operation: WP1 and WP2}\label{wp1}

\begin{figure*}
	\begin{center}
		\includegraphics[width=1\textwidth]{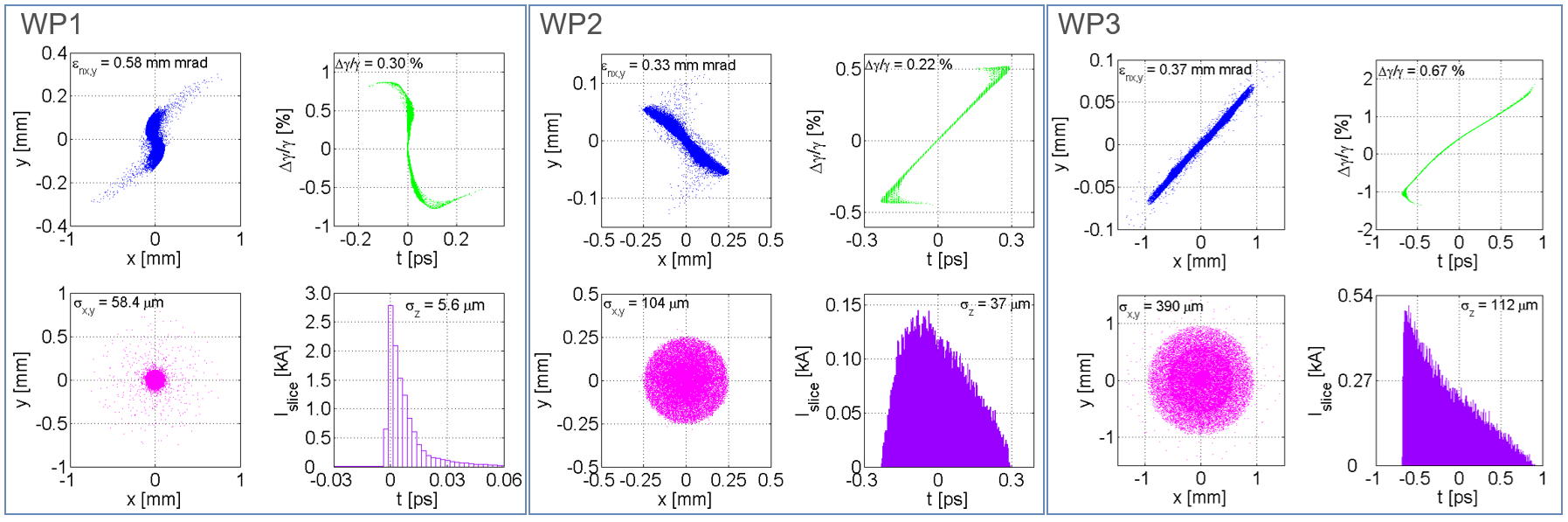}    
		\caption{Transverse (x and y) and longitudinal phase spaces (upper plots) and transverse distribution and current profile (lower plots) for WP1, WP2 and WP3. The results are output from TStep code at the photo-injector.}
		\label{fig:inj_phsp}
	\end{center}
\end{figure*}


Detailed beam dynamics studies have been performed to drive a witness bunch suitable both for PWFA and LWFA. The case of interest foresees at the undulator entrance a 0.5 GeV witness beam energy with less than 1 mm mrad emittance and 30 pC in 10 fs FWHM length, that means 3 kA peak current. Two different configurations have been studied for the photo-injector: WP1 foresees a fully RF compression applying the velocity bunching scheme (more details in \cite{item303}) and WP2 a hybrid compression (at least a factor 3 from the photo-injector) relying in an RF compression stage followed by a magnetic one. In both cases the choice has been to explore the blow-out regime by adopting a photo-cathode laser with a gaussian longitudinal profile of RMS duration $\sigma_z$ $\simeq$ 120 $\mu m$ and a transverse uniform distribution of RMS spot size $\sigma_r$ = 175 $\mu m$. 

Beam dynamics simulations result is a 30 pC bunch exiting from the photo-injector with RMS transverse normalised emittance $\epsilon_{n_{x,y}}$ $<$ 0.6 mm-mrad and RMS bunch length, $\sigma_z$ tunable in the range 5 - 40 $\mu m$ and are shown in  \figurename~\ref{fig:inj_phsp} and \figurename~\ref{fig:inj_phsp}: the longitudinal and transverse phase spaces are reported at the photo-injector exit as obtained with Tstep. It is worth to notice that the spike-like current distribution with the charge gathered on the head of the bunch, typical of the velocity bunching regime, is suitable in order to take profit of the beam loading in the plasma acceleration stage.

\subsection{Photo-injector sensitivity studies}\label{sens_ph}
Photo-injector sensitivity studies have been performed in terms of the exiting beam quality to test the robustness of the adopted working point. The analysis has been carried out looking at WP1 that is the most sensitive to the photo-injector phase stability, needed to ensure 3 $\mu m$ FWHM witness length. Machine sensitivity studies have been performed on samples of 80 machine runs. For each machine run has been generated a Tstep input code in which the applied errors have been randomly generated inside the chosen ranges reported in \tablename~\ref{tab:table_err}.

Taking advantage of the  experience acquired at SPARC$\_$LAB, the maximum reasonable error values have been considered to face the most realistic situation, trying not to count only on the best performance of the machine systems. Since the RF compression occurs in the accelerating cavities and it strongly depends on the RF phase stability, at first phase and voltage jitter on the accelerating cavities have been introduced one by one with the aim to determine most dangerous error contributions. Then jitter on the RF gun power system and on the extracted beam charge have been introduced. 

\begin{table}
	\small
	\begin{center}
		\caption{Studied jitter for the RF gun, accelerating cavities and photocathode laser system}
		\label{tab:table_err}
		\begin{tabular}{|l|c|c|}
			\hline
			GUN  && \\
			\hline
			RF Voltage [$\Delta V_G$]   & \centering $\pm$ 0.2 & $\%$  \\
			RF Phase [$\Delta \Phi_G$] & \centering $\pm$ 0.05,  $\pm$ 0.1 & $deg$ \\
			\hline
			S-band Accelerating Sections&&\\
			\hline
			RF Voltage [$\Delta V_S$] & \centering $\pm$ 0.2& \% \\
			RF Phase [$\Delta \Phi_S$]& \centering  $\pm$0.05,  $\pm$ 0.1&  $deg$ \\
			\hline
			Cathode Laser System&&\\
			\hline
			Charge Fluctuation [Q] & \centering $\pm$ 5 & \% \\
			\hline
		\end{tabular}
	\end{center}
\end{table}

The data analysis highlights the effect of the RF phase jitter on the beam length, while other jitter contributions do not cause significant degradation of the beam parameters.
Less than 5 $\%$ machine runs goes in peak current lower than the required 3 $\mu m$ in the FWHM beam length and they correspond to absolute RF phase jitter higher than 0.05 degree. All the parameters are still compliant with the required ones and the analysis suggests that RF phase jitter lower than $\pm$ 0.05 degree can ensure the needed beam peak current. The results overall the 80 machine runs are summarised in \tablename~\ref{tab:30_err}.

\begin{table}
	\small
	\begin{center}
		\caption{Simulated parameters for the 30 pC electron beam at the photo-injector exit in case of any errors and in case of jitter on the machine as described in \tablename~\ref{tab:table_err}.}	
		\label{tab:30_err}		
		\begin{tabular}{|l|c|c|c|}
			\hline
			\textbf{@Ph. Exit}		   & Without errors		& \centering With errors	 &  \\
			\hline
			E	 &98.8					   &\centering 98.8    $\pm$ 0.5  & MeV \\
			\hline
			$\Delta\gamma/\gamma$							 &0.30		  				 &\centering 0.30   $\pm$ 0.01& $\%$\\ 
			\hline
			$\epsilon_{n_{x,y}}$							   &0.58		  			  &\centering 0.58    $\pm$ 0.02  & mm mrad \\
			\hline
			$\sigma_z$							  &5.6		  			&\centering 5.6 $\pm$ 0.1    & $\mu m$ \\
			\hline
			$\tau_z$							        &3		  				&\centering 3   $\pm$ 0.2    & $\mu m$  \\
			\hline
			$I_{peak}$	         &3	  				        &\centering 3     $\pm$  0.5     & $\mu m$  \\ 
			\hline

		\end{tabular}
	\end{center}
\end{table}

\section{High charge-Low Current from photo-injector: WP3}
A 200 pC electron beam with 30 k macro particles has been carried up to the X-band linac entrance to obtain the following beam features: peak current $I_{{peak}}$ $\geq$ 0.3 kA, bunch length $\sigma_z$ $\leq$ 120 $\mu m$ and transverse normalised emittance $\epsilon_{n_{x,y}}$ $\leq$ 0.5 mm mrad. The aim is to produce an electron beam with high phase space density and high brightness suitable for driving respectively Compton back-scattering radiation sources and FEL's machines.

An extensive simulation campaign, leads to consider a photo-cathode laser pulse with a gaussian longitudinal profile of 0.9 ps RMS duration and a transverse uniform distribution of spot size $\sigma_r$ = 170 $\mu m$, radius r = 0.35 mm. The photo-injector is operated in the velocity bunching scheme in the first S-band cavity to shorten the RMS beam length from 270 to $\simeq$ 112 $\mu m$. In this configuration the parameters of the electron beam exiting the photo-injector are: E = 171.4 MeV, $\epsilon_{n_{x,y}}$ = 0.37 mm mrad, transverse spot size $\sigma_{x,y}$ = 390 $\mu m$, $\Delta\gamma/\gamma$ = 0.67 \%, $\sigma_z$ = 112 $\mu m$. The longitudinal and transverse phase spaces are reported at the photo-injector exit in \figurename~\ref{fig:inj_phsp} as obtained with Tstep.

\section{Particle driven configuration: WP4}
A \textquoteleft comb-like\textquoteright ~configuration for the electron beam, consisting of a 200 pC driver followed by a 30 pC witness bunch generated through the so-called \textquoteleft laser-comb technique\textquoteright ~ \cite{item50}, has been explored aiming to optimise the witness parameters and to set the longitudinal distance between the two bunches at a desired value. 
Such operating mode enables the possibility to pilot a PWFA stage where the passage of an ultra-relativistic bunch of charged particles (the driver) through a plasma drives a charge density wake useful to accelerate the trailing bunch.

Computational studies have been dedicated to provide at the plasma two bunches, i.e. driver and witness, separated by at least 0.55 ps, which corresponds to $\lambda_p$/2, being the plasma wavelength $\lambda_p$ $\simeq$ 334 $\mu m$ for a plasma background density $n_p$ = $10^{16}$ $cm^{-3}$. Both driver and witness bunches must be compressed down to $\simeq$ 50 fs and 10 fs (FWHM), respectively: the witness bunch length must be much less than the plasma wavelength in order to minimize the energy spread growth. In addition, one more request is on the minimization of the emittance growth, which unavoidably occurs because of the witness-driver overlapping.

\begin{figure}
	\begin{center}
		\includegraphics[width=0.8\columnwidth]{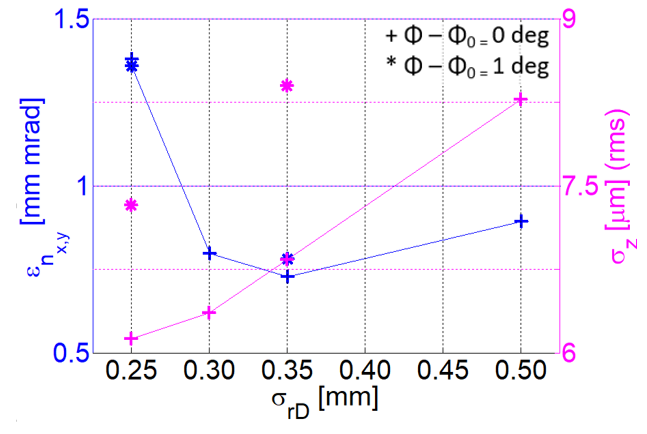}    
		\caption{Witness RMS transverse normalised emittance and RMS length at the photo-injector exit as function of the driver spot size at the cathode (blue crosses are for nominal 2nd TW cavity RF phase, while magenta stars are for RF phase increased of 1 degree with respect to the nominal one.}
		\label{fig:scan_driver1}
	\end{center}
\end{figure}

\begin{figure}
	\begin{center}
		\includegraphics[width=0.95\columnwidth]{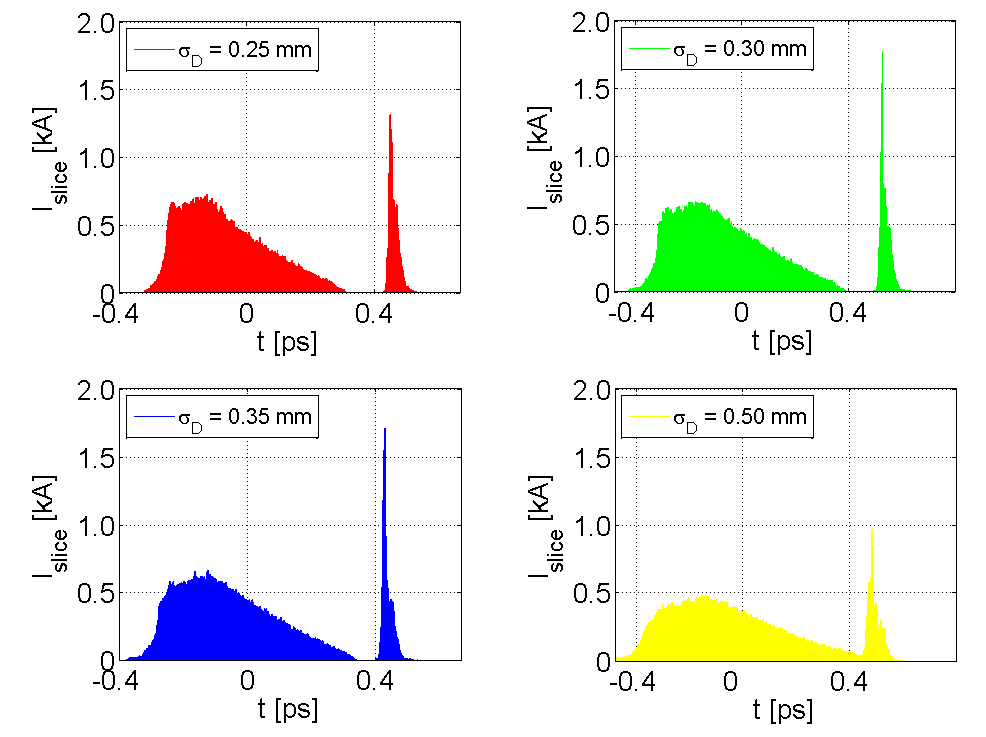}    
		\caption{Longitudinal distribution of the comb-beam at the photo-injector exit for several driver spot size at the cathode. The beam is propagating from right to left with the driver arriving earlier that the witness. }
		\label{fig:scan_driver2}
	\end{center}
\end{figure}

The photo-cathode laser has been shaped in order to provide at the cathode the witness pulse, whose distribution is described in \ref{wp1}, separated by $\simeq$ 4 ps from the driver pulse. A gaussian longitudinal distribution with $\sigma_z$ = 120 $\mu m$ and a uniform transverse distribution of $\sigma_{rD}$ = 350 $\mu m$ has been assumed for the driver. The driver spot size on the cathode has been chosen looking at the witness quality, the witness emittance and longitudinal profile depending on it. The behaviour of the witness transverse normalised emittance as function of the driver spot size indicates $\sigma_{rD}$ = 350 $\mu m$ as the optimal value for the driver spot size at the cathode surface, as shown in the plot in \figurename~\ref{fig:scan_driver1}: once chosen the accelerator set-up, is possible to tune the phase space densities of the crossing beams to minimise the witness quality degradation. In addition, it is worth to notice that by adopting a $\sigma_{rD}$ = 350 $\mu m$ the FWHM witness length does not suffer lengthening, as shown in \figurename~\ref{fig:scan_driver2}, although the lower RMS witness length is obtained for $\sigma_{rD}$ = 250 $\mu m$.

Besides of an appropriate shaping and relative spacing of the laser-comb pulses at the cathode surface, a proper set of active and passive accelerator elements allows us to obtain the required comb beam at the photo-injector exit. The choice of the accelerator set-up starts from the optimised witness working point illustrated in \ref{wp1}, with additional fine-tuning of accelerating cavity RF phases and solenoid magnetic fields.


\begin{figure}
	\begin{center}
		\includegraphics[width=0.95\columnwidth]{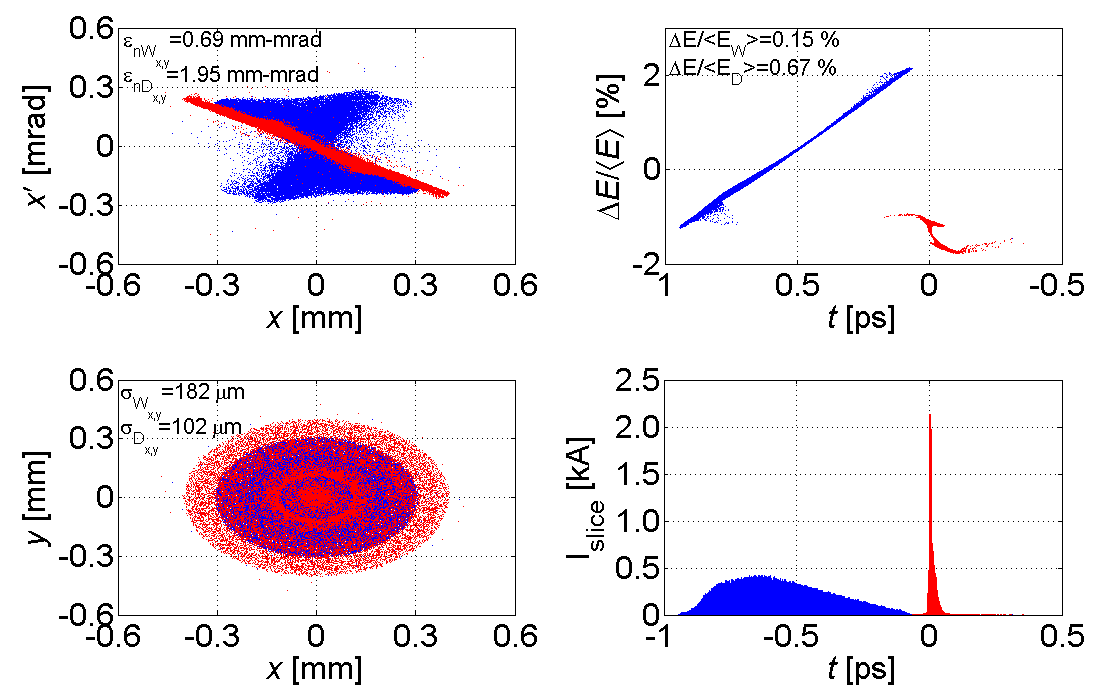}    
		\caption{Upper plots: transverse and longitudinal phase spaces. Lower plots: transverse distribution and current profile. The results are output from TStep code at the photo-injector exit in case of comb-like operation. The blue and red dots are related to the driver and witness respectively. }
		\label{fig:inj_phsp_comb}
	\end{center}
\end{figure}

Best compromise in terms of final spacing and witness profile has been obtained with a laser-comb operation with two laser pulses spaced of $\Delta t$ = 4.8 ps on the cathode. In this configuration, adopting the set-up described in \ref{wp1}, the beam crossing occurs in the second TW accelerating cavity and a fine-tuning of the RF phases suffices to provide $\simeq$~0.55 ps spaced beams and the desired witness and driver longitudinal lengths: 3 $\mu m$ FWHM and in the range 30 - 50 $\mu m$ RMS,  respectively.
Both witness and driver bunches have been simulated with 30k and 200k
macro-particles respectively, corresponding to 30 pC and 200 pC. In the described configuration the driver arrives 0.58 ps earlier than the witness at the X-band booster. The parameters of witness and driver at the X-band linac entrance are listed in \tablename~\ref{tab:table_inj}: it is worth to notice that the FWHM witness length is of about 3 $\mu m$, and so the peak current of $\simeq$ 3 kA, nevertheless the $\epsilon_{n_{X,Y}}$ grows up to 0.7 mm-mrad. \figurename~\ref{fig:inj_phsp_comb} reports the 6D phase space at the photo-injector exit for both witness and driver beams as obtained with Tstep.

\section{Conclusions}
Electron beam dynamics in the SPARC like photo-injector has been detailed studied for several working points, each one corresponding to a case of interest for the EuPRAXIA@SPARC$\_$LAB facility. Promising results have been obtained in terms of phase space quality for the electron beam enabling the generation of high brightness – ultra-short fs bunches with up to 3 kA peak current and a transverse normalised emittance lower than 0.8 mm-mrad, whatever working point is considered. Sensitivity studies have been also presented for WP1, the most sensitive to the photo-injector stability, showing the machine design robustness inside the tolerance range chosen on the base of routine operation at the SPARC$\_$LAB facility. In the next future sensitivity studies will be enlarged to the WP2, WP3 and WP4. 

\section*{Acknowledgment}
This work was supported by the European Union's Horizon 2020 research and innovation programme under grant agreement No. 653782.






\end{document}